
\input harvmac
\input epsf
\noblackbox
\message{S-Tables Macro v1.0, ACS, TAMU (RANHELP@VENUS.TAMU.EDU)}
\newhelp\stablestylehelp{You must choose a style between 0 and 3.}%
\newhelp\stablelinehelp{You should not use special hrules when stretching
a table.}%
\newhelp\stablesmultiplehelp{You have tried to place an S-Table inside another
S-Table.  I would recommend not going on.}%
\newdimen\stablesthinline
\stablesthinline=0.4pt
\newdimen\stablesthickline
\stablesthickline=1pt
\newif\ifstablesborderthin
\stablesborderthinfalse
\newif\ifstablesinternalthin
\stablesinternalthintrue
\newif\ifstablesomit
\newif\ifstablemode
\newif\ifstablesright
\stablesrightfalse
\newdimen\stablesbaselineskip
\newdimen\stableslineskip
\newdimen\stableslineskiplimit
\newcount\stablesmode
\newcount\stableslines
\newcount\stablestemp
\stablestemp=3
\newcount\stablescount
\stablescount=0
\newcount\stableslinet
\stableslinet=0
\newcount\stablestyle
\stablestyle=0
\def\stablesleft{\quad\hfil}%
\def\stablesright{\hfil\quad}%
\catcode`\|=\active%
\newcount\stablestrutsize
\newbox\stablestrutbox
\setbox\stablestrutbox=\hbox{\vrule height10pt depth5pt width0pt}
\def\stablestrut{\relax\ifmmode%
                         \copy\stablestrutbox%
                       \else%
                         \unhcopy\stablestrutbox%
                       \fi}%
\newdimen\stablesborderwidth
\newdimen\stablesinternalwidth
\newdimen\stablesdummy
\newcount\stablesdummyc
\newif\ifstablesin
\stablesinfalse
\def\begintable{\stablestart%
  \stablemodetrue%
  \stablesadj%
  \halign%
  \stablesdef}%
\def\stablesadj{%
  \ifcase\stablestyle%
    \hbox to \hsize\bgroup\hss\vbox\bgroup%
  \or%
    \hbox to \hsize\bgroup\vbox\bgroup%
  \or%
    \hbox to \hsize\bgroup\hss\vbox\bgroup%
  \or%
    \hbox\bgroup\vbox\bgroup%
  \else%
    \errhelp=\stablestylehelp%
    \errmessage{Invalid style selected, using default}%
    \hbox to \hsize\bgroup\hss\vbox\bgroup%
  \fi}%
\def\stablesend{\egroup%
  \ifcase\stablestyle%
    \hss\egroup%
  \or%
    \hss\egroup%
  \or%
    \egroup%
  \or%
    \egroup%
  \else%
    \hss\egroup%
  \fi}%
\def\stablestart{%
  \ifstablesin%
    \errhelp=\stablesmultiplehelp%
    \errmessage{An S-Table cannot be placed within an S-Table!}%
  \fi
  \global\stablesintrue%
  \global\advance\stablescount by 1%
  \message{<S-Tables Generating Table \number\stablescount}%
  \begingroup%
  \stablestrutsize=\ht\stablestrutbox%
  \advance\stablestrutsize by \dp\stablestrutbox%
  \ifstablesborderthin%
    \stablesborderwidth=\stablesthinline%
  \else%
    \stablesborderwidth=\stablesthickline%
  \fi%
  \ifstablesinternalthin%
    \stablesinternalwidth=\stablesthinline%
  \else%
    \stablesinternalwidth=\stablesthickline%
  \fi%
  \tabskip=0pt%
  \stablesbaselineskip=\baselineskip%
  \stableslineskip=\lineskip%
  \stableslineskiplimit=\lineskiplimit%
  \offinterlineskip%
  \def\borderrule{\vrule width \stablesborderwidth}%
  \def\internalrule{\vrule width \stablesinternalwidth}%
  \def\thinline{\noalign{\hrule height \stablesthinline}}%
  \def\thickline{\noalign{\hrule height \stablesthickline}}%
  \def\trule{\omit\leaders\hrule height \stablesthinline\hfill}%
  \def\ttrule{\omit\leaders\hrule height \stablesthickline\hfill}%
  \def\tttrule##1{\omit\leaders\hrule height ##1\hfill}%
  \def\stablesel{&\omit\global\stablesmode=0%
    \global\advance\stableslines by 1\borderrule\hfil\cr}%
  \def\el{\stablesel&}%
  \def\elt{\stablesel\thinline&}%
  \def\eltt{\stablesel\thickline&}%
  \def\elttt##1{\stablesel\noalign{\hrule height ##1}&}%
  \def\elspec{&\omit\hfil\borderrule\cr\omit\borderrule&%
              \ifstablemode%
              \else%
                \errhelp=\stablelinehelp%
                \errmessage{Special ruling will not display properly}%
              \fi}%
  \def\stmultispan##1{\mscount=##1 \loop\ifnum\mscount>3 \stspan\repeat}%
  \def\stspan{\span\omit \advance\mscount by -1}%
  \def\multicolumn##1{\omit\multiply\stablestemp by ##1%
     \stmultispan{\stablestemp}%
     \advance\stablesmode by ##1%
     \advance\stablesmode by -1%
     \stablestemp=3}%
  \def\multirow##1{\stablesdummyc=##1\parindent=0pt\setbox0\hbox\bgroup%
    \aftergroup\emultirow\let\temp=}
  \def\emultirow{\setbox1\vbox to\stablesdummyc\stablestrutsize%
    {\hsize\wd0\vfil\box0\vfil}%
    \ht1=\ht\stablestrutbox%
    \dp1=\dp\stablestrutbox%
    \box1}%
  \def\stpar##1{\vtop\bgroup\hsize ##1%
     \baselineskip=\stablesbaselineskip%
     \lineskip=\stableslineskip%
     \lineskiplimit=\stableslineskiplimit\bgroup\aftergroup\estpar\let\temp=}%
  \def\estpar{\vskip 6pt\egroup}%
  \def\stparrow##1##2{\stablesdummy=##2%
     \setbox0=\vtop to ##1\stablestrutsize\bgroup%
     \hsize\stablesdummy%
     \baselineskip=\stablesbaselineskip%
     \lineskip=\stableslineskip%
     \lineskiplimit=\stableslineskiplimit%
     \bgroup\vfil\aftergroup\estparrow%
     \let\temp=}%
  \def\estparrow{\vfil\egroup%
     \ht0=\ht\stablestrutbox%
     \dp0=\dp\stablestrutbox%
     \wd0=\stablesdummy%
     \box0}%
  \def|{\global\advance\stablesmode by 1&&&}%
  \def\|{\global\advance\stablesmode by 1&\omit\vrule width 0pt%
         \hfil&&}%
  \def\vt{\global\advance\stablesmode by 1&\omit\vrule width \stablesthinline%
          \hfil&&}%
  \def\vtt{\global\advance\stablesmode by 1&\omit\vrule width
\stablesthickline%
          \hfil&&}%
  \def\vttt##1{\global\advance\stablesmode by 1&\omit\vrule width ##1%
          \hfil&&}%
  \def\vtr{\global\advance\stablesmode by 1&\omit\hfil\vrule width%
           \stablesthinline&&}%
  \def\vttr{\global\advance\stablesmode by 1&\omit\hfil\vrule width%
            \stablesthickline&&}%
  \def\vtttr##1{\global\advance\stablesmode by 1&\omit\hfil\vrule width ##1&&}%
  \stableslines=0%
  \stablesomitfalse}
\def\stablesdef{\bgroup\stablestrut\borderrule##\tabskip=0pt plus 1fil%
  &\stablesleft##\stablesright%
  &##\ifstablesright\hfill\fi\internalrule\ifstablesright\else\hfill\fi%
  \tabskip 0pt&&##\hfil\tabskip=0pt plus 1fil%
  &\stablesleft##\stablesright%
  &##\ifstablesright\hfill\fi\internalrule\ifstablesright\else\hfill\fi%
  \tabskip=0pt\cr%
  \ifstablesborderthin%
    \thinline%
  \else%
    \thickline%
  \fi&%
}%
\def\endtable{\advance\stableslines by 1\advance\stablesmode by 1%
   \message{- Rows: \number\stableslines, Columns:  \number\stablesmode>}%
   \stablesel%
   \ifstablesborderthin%
     \thinline%
   \else%
     \thickline%
   \fi%
   \egroup\stablesend%
\endgroup%
\global\stablesinfalse}

\def\sni{\smallskip\noindent}
\def\mni{\medskip\noindent}
\def\ie{{\it i.e.}}
\def\nl{\hfil\break}

\def\npb{{ \sl Nucl. Phys. }}
\def\prc{{ \sl Phys. Rep. }}
\def\prd{{ \sl Phys. Rev. }}
\def\prl{{ \sl Phys. Rev. Lett. }}
\def\plb{{ \sl Phys. Lett. }}

\def\zpc{{ \sl Zeit. Phys. }}
\def\IR{\relax{\rm I\kern-.18em R}}
\def\undertext#1{\vtop{\hbox{#1}\kern 1pt \hrule}}

\def\c#1{{\cal{#1}}}
\def\dirac{\hbox{$\partial$\kern-0.5em\raise0.3ex\hbox{/}}}
\def\dslash{\hbox{$\partial$\kern-0.5em\raise0.3ex\hbox{/}}}
\def\pslash{\hbox{{\it p}\kern-0.5em\raise-0.3ex\hbox{/}}}
\def\gsim{\mathrel{\raise.3ex\hbox{$>$\kern-.75em\lower1ex\hbox{$\sim$}}}}
\def\lesssim{\mathrel{\raise.3ex\hbox{$<$\kern-.75em\lower1ex\hbox{$\sim$}}}}
\def\d{\partial}

\def\gev{{\,\rm GeV}}
\def\tev{{\,\rm TeV}}

\def\muz{\mu}
\def\y{y_t}
\def\nf{N_F}
\def\nc{N_c}

\baselineskip=12pt
\Title{\vbox{\baselineskip12pt\hbox{TIT/HEP--258}\hbox{CERN-TH.7287/94}}}
{\titlefont{Triviality, perturbation theory and $Z\to b \overline b$}}
\centerline{Kenichiro Aoki\footnote{$^1$}
{internet:{\tt~ken@phys.titech.ac.jp}}
and Santiago Peris\footnote{$^2$}{On leave from Grup de Fisica Teorica and
IFAE, Universitat Autonoma de Barcelona, Barcelona, Spain.
internet:{\tt~peris@surya11.cern.ch}}}
\bigskip\centerline{\it ${}^1$Department of Physics}
\centerline{\it Tokyo Institute of Technology}
\centerline{\it Oh--okayama, Meguro-ku}
\centerline{\it Tokyo, JAPAN   152}
\bigskip\centerline{\it ${}^2$Theory Division, CERN,}
\centerline{{\it CH}{\rm --} 1211\  \it Geneva {\oldstyle 23},
Switzerland.}
\vskip .3in
\centerline{\bf Abstract}\nl
We compute the large-$m_t$ contributions to the
$Zb\overline b$ vertex in the Standard Model
non-perturbatively using both the $1/\nc$ and the $1/\nf$
expansions. Triviality and its consequences for the perturbative
result are systematically analyzed.
In particular, we point out that the effect of triviality
is already important at the two-loop level for a heavy Higgs
($600\gev<M_H<1\tev$) even when the top mass is below $200\gev$.

\vskip 1.5in

\leftline{June 1994}

\Date{CERN-TH.7287/94}

\subsec{Introduction }
A peculiar feature of the Standard Model (SM) is the appearance of
non-decoupling effects in radiative corrections.
This means that certain  contributions from heavy particles do not decrease
(and sometimes even grow) as their mass increases.
The most striking examples of this are the contribution of the top
quark and the Higgs boson to the $\rho$ parameter \ref\V{M. Veltman, \npb {\bf
B123} (1977) 89; Acta Phys. Pol. {\bf B8} (1977) 475; \plb {\bf B70} (1977)
253;
for a recent review on electroweak radiative corrections see, for instance, F.
Jegerlehner, TASI 1990, Testing the Standard Model, proceedings, edited by
M. Cvetic and P. Langacker, World Scientific, Singapore, 1991.}  and
the $Zb\overline b$ vertex \ref\zbb{A. Akhundov, D. Bardin and T. Riemann, \npb
{\bf
B276} (1986) 1; W. Beenakker and W. Hollik\zpc {\bf C40} (1988) 141; J.
Bernabeu, A. Pich and A. Santamaria, \plb {\bf B200} (1988) 569,\npb {\bf
B363} (1991) 326. For a review, see also F. Jegerlehner in ref. \V .}.
Moreover, we are presently witnessing the test of SM radiative
corrections against
high-precision experiments, where impressive accuracies
are being attained \ref\LEP{R. Miquel, preprint CERN-PPE/94-70. Talk
given at the
22nd INS Symposium on "Physics with High Energy Colliders", Tokyo, Japan, March
1994.}.
On the theoretical side, this experimental accomplishment has
spurred the quest for  more precise calculations with a
better control over higher-order contributions. To this end,
two-loop electroweak effects  are now being obtained and analyzed in a rather
systematic way \ref\twoloop{R. Barbieri, M. Beccaria, P. Ciafaloni, G. Curci
and A. Vicere, \plb {\bf B288} (1992) 95, Err. {\bf B312} (1993) 511, \npb
{\bf B409} (1993) 105; J. Fleischer, O.V. Tarasov and F. Jegerlehner, \plb
{\bf B319} (1993) 249; G. Degrassi, S. Fanchiotti and P. Gambino,
preprint CERN-TH-7180-94, hep-ph 9403250. For earlier work see
J.J. van der Bij and
F. Hoogeveen, \npb {\bf B283} (1987) 477 and J.J. van der Bij and M.
Veltman, \npb {\bf B231} (1984) 205.}
. Given this state of affairs,
we think it is important to understand the behavior of the
perturbative series itself and
assess its validity and limitations.
With this idea in mind, we previously studied
the dependence of the $\rho$ parameter on the top and the Higgs mass
non-perturbatively using a $1/N$ expansion \ref\AP{
K.~Aoki and S. Peris, \zpc{\bf C61} (1994) 303. See also S. Peris, \plb {\bf
B251} (1990) 603.}.
In the present letter, we shall analyze the leading
contribution of the top quark to the $Zb\overline b$ vertex.
This is of interest since the $Zb\overline b$ vertex, besides the $\rho$
parameter, is the only other independent example where an
$m_t^2$ growth is seen in perturbation theory.

The $1/N$ approach is a systematic and controlled expansion
of the Schwinger--Dyson equations
and is  non-perturbative in that it resums an
infinite set of Feynman diagrams \ref\LARGEN{See, for instance,
S.~Coleman, in {\sl ``Aspects of symmetry"}, Cambridge University
Press, Cambridge, 1985.}.
Consequently, we can analyze certain
non-perturbative properties, such as triviality, that cannot be treated
in a perturbative approach.
There are a variety of analyses (most noticeably the
lattice) leading to the result that the SM is a trivial theory
\ref\TRIVIALITY{See for instance D.J.E.~Callaway, \prc{\bf167} (1988) 241
and references therein; for a recent
analysis see W. Bock, C. Frick, J. Smit and J.C. Vink, \npb{\bf B400} (1993)
309.}. Triviality, succinctly, means that the SM has a  physical scale
$\Lambda_{triv}$ beyond which the theory ceases to make sense
so that it
becomes an effective theory valid {\it only} for momentum scales
smaller than $\Lambda_{triv}$ \ref\CONSOLI{This is the common lore. For an
alternative viewpoint (as far as the pure scalar sector is concerned)
see however M. Consoli and P.M. Stevenson, hep-ph 9403299 and references
therein.}.
Using the vertex $Zb\overline b$ as an
example,
our objective here will be
to systematically analyze how the finite energy scale
$\Lambda_{triv}$ appears in physical quantities in the SM
and to assess its consequences for perturbation theory.

Commonly,
radiative effects are studied
using perturbation theory. While this is indeed a reasonable approximation
for
small couplings, non-decoupling effects become more appreciable  when
these coupling constants are larger and,
in an extreme case, even leave the perturbative regime.
Perturbation theory leads one to believe that, as these couplings grow
within the perturbative domain, more and more loops are to be computed
to attain the desired degree of precision in the calculation. The result we
find is that for large masses of the Higgs (or the top) this
is not  true.
 Because of non-perturbative effects, namely triviality and the
corresponding appearance of a physical cutoff, when the Higgs mass
becomes  heavy (but still lighter than a TeV, say) cutoff effects
introduce an intrinsic uncertainty in the
perturbative expression that grows very fast with the Higgs mass. Similar
conclusions were reached in ref. \ref\PP{S. Cortese, E. Pallante and R.
Petronzio, \plb {\bf B301} (1993) 203.}.
This ambiguity can be a sizeable fraction of the two-loop top
contribution to the $Zb\overline b$ vertex if $M_H\gsim 600\gev$, even when
the top mass is smaller than $200\gev$, for instance.
At this point the calculation
of higher orders does not improve the precision of the results.
Clearly, the situation would change if the underlying theory for $\mu\gsim
\Lambda_{triv}$ were known since, in principle, these
cutoff effects would then be
precisely calculable.

To consistently study how the triviality of the Higgs system
afflicts physical quantities,
one clearly needs to go beyond perturbation
theory since
otherwise  there is no triviality at all. Here, we shall use a
large-$N$ approach
because it is a very convenient way to implement triviality in the dynamics
and still keep the treatment analytical.
Using the fact that the Yukawa sector is by itself also trivial
\ref\FLAT{J.~Shigemitsu, \plb{\bf226B} (1989) 364;
I-H.~Lee, J.~Shigemitsu and R.~Shrock, \npb{\bf B330} (1990) 225,
{\bf B335} (1990) 265;
W.~Bock, A.K.~De, C.~Frick, K.~Jansen and T.~Trappenburg,
\npb{\bf B371}~(1992)~683;
S.~Aoki, J.~Shigemitsu and J.~Sloan, \npb{\bf B372} (1992) 361;
Proceedings of the International Symposium on Lattice Field Theory,
KEK (1991), and references therein. See also W. Bock et al. in ref.
\TRIVIALITY. }%
\ref\EG{M.B.~Einhorn and G.~Goldberg, \prl{\bf57 }(1986) 2115.}
\ref\KA{K. Aoki, \prd{\bf D44} (1991) 1547.}\ref\BN{G. Bathas and H.
Neuberger\plb{\bf B293} (1992) 417.}, we shall
consider  non-perturbative large-$N$ calculations (for $\nc$ and $\nf$)
of the $Zb\overline b$ vertex in order to gain some understanding of the
interplay between
triviality and perturbation theory. This will allow us to
estimate the effects of triviality
for the more interesting case of a heavy Higgs.
On the more theoretical side, we are also interested in finding out
whether the growth with the top mass seen in perturbation theory in
the $Z\overline b b$ vertex is bounded due to
non-perturbative effects, as happens with the $\rho$ parameter \AP. As we
shall see, this is indeed the case.

The  first observation is that leading $m_t$
contributions are entirely
due to the dynamics of the Yukawa sector of the SM \ref\ng{R.S.
Lytel, \prd {\bf
D22} (1980) 505; S. Peris in ref. \AP; R. Barbieri et al. in
ref. \twoloop .};
they, in
essence, have nothing to
do with the gauge field dynamics even though we always measure them in
the theory once it is ``gauged". This is also true for the $Zb\overline b$
vertex, as we will now briefly review.
The starting SM Lagrangian  (with $g=g'=0$) reads
\eqn\one{-{\cal L}=\overline \psi_L  \dslash \psi_L +\overline
t_{R}  \dslash t_{R} +y_t( \overline \psi_L \phi t_R +   {\rm h.c.}) - {\cal
L}(\phi)\quad ,}
with $\phi^T=\left({(v+H+i\chi^0)/\sqrt{2}},\ -\chi^-\right)$;
$\chi^{\pm,0}$ being the
Nambu--Goldstone bosons and ${\cal L}(\phi)$ the usual kinetic term
and the potential that yields a vacuum expectation value $v/\sqrt2$
for the scalar field $\phi$; $\psi_L$ is the top-bottom
left-handed doublet,
and $t_{\scriptscriptstyle R}$ the right-handed top singlet.
We consider  the  bottom quark as massless whereas the top has a mass
$m_t=y_t v/\sqrt{2}$ at tree level.
Quantum corrections induce an effective derivative coupling between
the Nambu--Goldstone boson and the $b$ quark that is given by
\eqn\two{-{\cal L}= \tau(y_t) \overline b_L \gamma^{\mu} b_L \partial_{\mu}
{\chi^0\over v}+\overline b \dslash b+\ldots}
When one gauges the theory by the usual replacement $\partial_{\mu}
\to D_{\mu}$ it leads to the following interactions with the $Z^0$,
\eqn\three{-{\cal L}= \tau(y_t) \overline b_L \gamma^{\mu} b_L
\left(\partial_{\mu}
{\chi^0\over v}-{g\over 2c} Z_{\mu}\right) + {g\over c} \overline b
\gamma^{\mu}
\left(-{1\over 2} P_L +{1\over 3} s^2\right)b Z_{\mu}+\ldots}
so that, in the unitary gauge, one obtains
\eqn\four{-{\cal L}= {g\over c} \overline b \gamma^{\mu}
[g_V+g_A \gamma_5]b
Z_{\mu}+\ldots}
with
\eqn\five{g_V=-{(1+\tau(\y))\over 4}+{s^2\over 3}\qquad ;\qquad
g_A={(1+\tau(\y))\over 4}\quad .}
Here, the notation $c\equiv\cos\theta_W, s\equiv\sin\theta_W$ is
used.
A non-vanishing value of $\tau(y_t)$ modifies the ratio of vector
to axial couplings of the bottom to the $Z^0$ with respect to the same
ratio for the lighter fermions. This can be measured, for instance, by
comparing forward--backward asymmetries for the bottom quark and for the
other fermions \ref\asym{See for instance M. Bohm and W. Hollik in "Z.
Physics at LEP 1", edited by G. Altarelli, R. Kleiss and C. Verzegnassi, CERN
report 89-08; F. Boudjema, A. Djouadi and C. Verzegnassi\plb {\bf
B238} (1990) 423.}.
In perturbation theory, the leading-order contribution in the
expansion with respect to $y_t^2\equiv 2\sqrt{2}G_Fm_t^2$ is
\eqn\five{\tau(y_t)= - {\sqrt{2} G_F m^2_t\over 8 \pi^2}\qquad . }
\subsec{Triviality and the $Zb\overline b$ vertex in the $1/\nc$
expansion}
We shall now
compute $\tau(\y)$ non-perturbatively in $\y^2$.
The $1/\nc$ expansion is performed by letting $\nc \to \infty$ while
taking $y_t^2\nc$ and
$v^2/\nc$ to be quantities of order one.
To leading order in $1/\nc$,
the top mass  is not corrected and the diagram of \fig\figtwo{}\
corrects the charged Nambu--Goldstone propagator.
\nl\centerline{\epsfysize=2.5cm\epsfbox{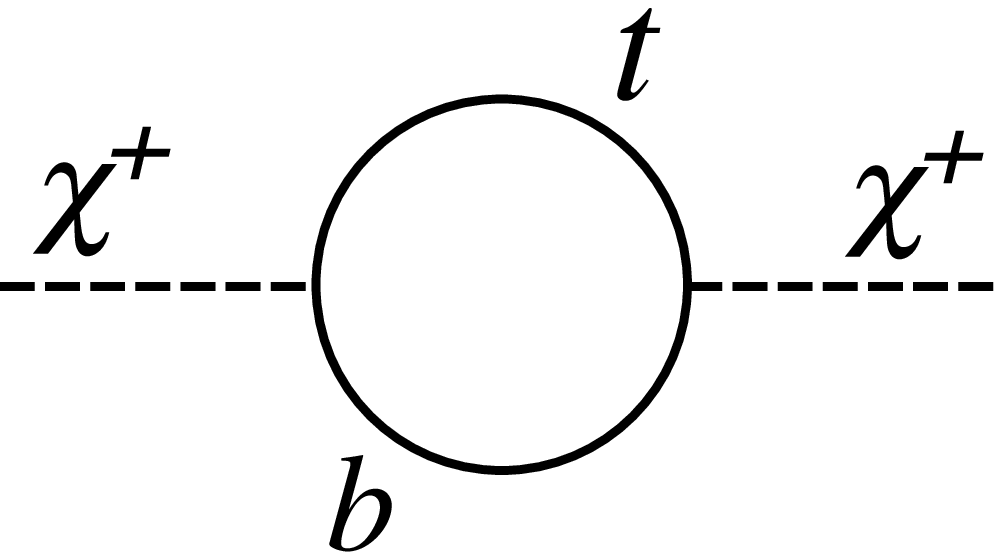}}\smallskip\noindent
\sni\figtwo\ {\it Leading-order correction to the charged
Nambu--Goldstone boson propagator in the $1/\nc$ expansion.}\sni

\vskip 0.25in
When this quantum
effect is taken into account, the effective Lagrangian for the charged
Nambu--Goldstone boson reads
\eqn\six{-{\cal L}= Z_+ (\d_{\mu} \chi^+- {gv\over 2} W^+_{\mu})^2
+\ldots \quad ,\quad\hbox{\rm where}\quad
Z_+=1-{\y^2 \nc\over (4\pi)^2} \log{m_t^2\over L^2}\quad ;}
$L$ may be thought of as a momentum cutoff\foot{$\log{{(L^2/
4\pi)}}= {2/( 4-n)}-\gamma_{\scriptscriptstyle E}+1/2$ in dimensional
regularization.}
; $W^+$ enters eq. \six\ in this precise combination because of
gauge invariance.
The presence of $Z_+$ implies that the $W^+$ mass term in the
Lagrangian is now $Z_+ g^2 v^2/4$ and consequently, the Fermi constant
appearing for instance in $\mu$ decay is
\eqn\seven{{\left(\sqrt2G_F\right)^{-1}}={ Z_+ v^2}\equiv V^2}
This equation expresses how the bare parameter $v$ is related to the
physical observable $G_F$.
This is all we need, for now the Yukawa coupling
$\y$ renormalizes (\ie\ it can be expressed in terms of physical
parameters) as
\eqn\eight{\y^2=2 {m_t^2\over v^2}
=2{(m_t^2)^{phys}\over v^2}=2\sqrt{2}
G_F(m_t^2)^{phys} Z_+\equiv (\y^R)^2 Z_+ \quad .}
This last equation may be reinterpreted as
a running coupling constant
\eqn\nine{\y^2(\mu')=
{\y^2(\mu)\over 1-{\y^2(\mu)\nc/ (4\pi)^2}\log
(\mu'^2/ \mu^2)}\quad ,}
with $\y\equiv \y(\mu'=L)$ and $\y^R\equiv \y(\mu=m_t)$.
It should be stressed that this large-$\nc$ equation is
non-perturbative in the Yukawa
coupling.\foot{Usual one-loop renormalization group arguments also lead
to an equation that looks like \nine, but with $\nc\to \nc + {3/ 2}$.}
Here,  triviality manifests itself through
the appearance of a scale $\Lambda_{triv}$ at which
$\y$ diverges.
This scale is to be interpreted as the maximum energy scale up
to which the theory, which now becomes an effective theory, is valid.
Equation \nine\ tells us that
\eqn\ten{\Lambda^2_{triv}=m^2_t \,exp\left({(4\pi)^2\over 2\sqrt{2} G_F
m^2_t \nc}\right)\quad ,}
which is a fantastically large scale for small values of $m_t$.
Imposing the condition that the top mass needs to be substantially
below the cutoff of the theory,
a triviality mass bound for the top may be obtained
 in the $1/\nc$  limit.
For instance, imposing $m_t/\Lambda_{triv}<0.4$, as is sometimes done,
we obtain $m_t<3.7\,V$, which is consistent with the
mass bound we obtain from the $1/\nf$ expansion, $m_t<3.5\,V$
\EG,\KA, from more sophisticated $1/N$ approaches \BN,  and also from
the lattice \FLAT.

Let us now turn to the actual calculation of $\tau(\y^R)$.
The leading $1/\nc$ contribution is given by \fig\figthree{}A,B.
\nl\centerline{\epsfysize=4.0cm\epsfbox{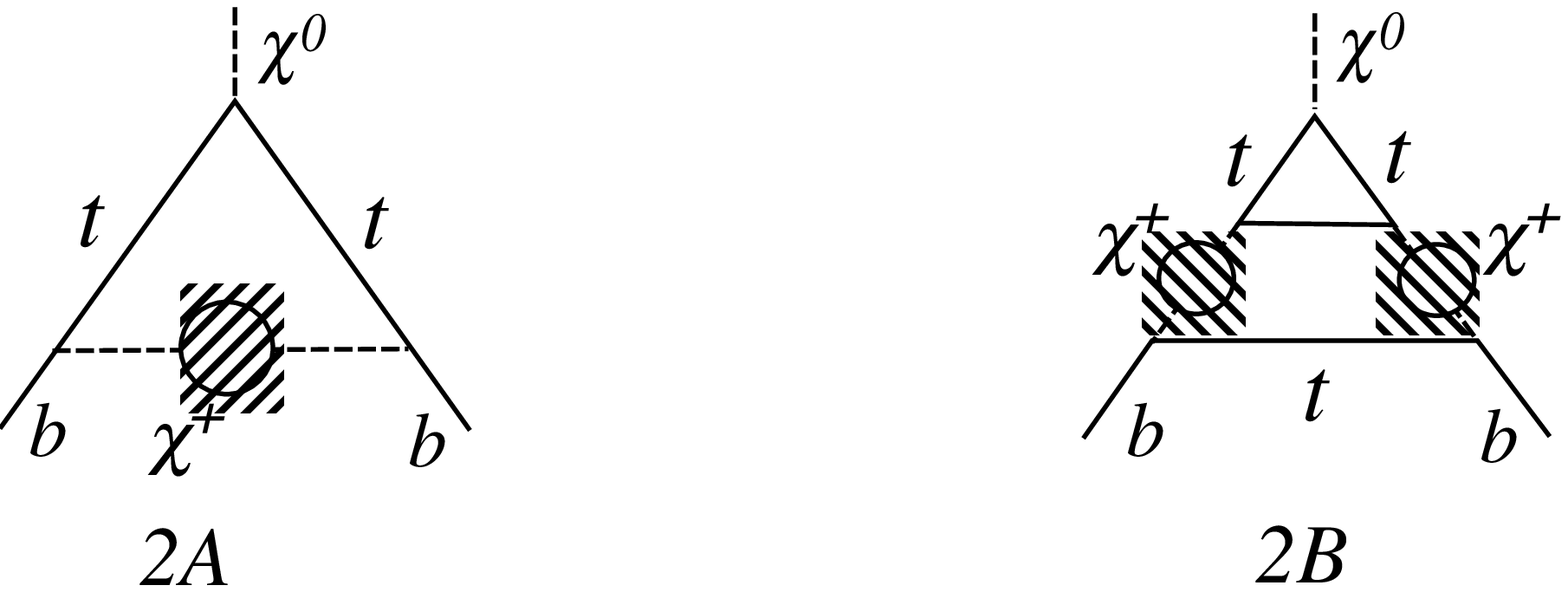}}
\sni \figthree A,B\ {\it  Leading-order contributions to
$\tau(\y^R)$ in $1/\nc$ expansion.
Lines with blobs denote the full propagators.}\sni

\vskip 0.25in
A straightforward
calculation, taking care of the due renormalization,  secures
\eqn\eleven{\eqalign{\tau(\y)&=\ \tau({\rm\figthree A}) +\
\tau({\rm\figthree B})\cr
\tau({\rm\figthree A})&=-\left({\y^R\over 4\pi}\right)^2
\int_0^{\infty}  {d\xi \over (1+\xi)^2 f(\xi)}\cr
\tau({\rm\figthree B})&= -{\nc\over 2}\left({\y^R\over
4\pi}\right)^4\int_0^{\infty} d\xi\
{1-(1+{1\over \xi}) \log(1+\xi)\over \xi\ (1+\xi)\  [f(\xi)]^2}\quad ,\cr}}
where
\eqn\thirteen{f(\xi)\equiv1-2 \nc \left({\y^R\over 4\pi}\right)^2
\int_0^1 dt\ t\  \log[1+\xi\ (1-t)]}
and $\xi=l_E^2/m_t^2$ for a Euclidean momentum $l_E$ routing inside
the loop\foot{In \figthree B the fermion loop connecting the three
Nambu--Goldstone bosons is finite.}.
Equations. \eleven\  are well behaved in the ultraviolet
($\xi\to \infty$) and in the infrared ($\xi\to 0$).
However there is a pole in the integrand
at a finite  momentum $\xi_0$.
This fact may  be traced back to the
appearance of denominators such as the one of eq. \nine\  and
is therefore a {\it direct} consequence of the triviality of the
Yukawa theory.
This is how triviality manifests itself in Green functions\foot{We
may also use the location of this pole as the definition of the
triviality scale. This would differ from $\Lambda_{triv}$ obtained
in \nine\ approximately by a factor of $e^{3/2}$. Which scale
we use will make no difference in our subsequent discussion.}.
If one expands eqs. \eleven\   in a power series in $\y^R$,
there is no pole and these expressions are well behaved. But, then,
there is no triviality in eq. \nine\  either!

Notice that, in fact, the pole is located at a momentum scale
that is beyond the triviality cutoff defined by eqs. \nine, \ten; i.e.
beyond the  physical cutoff up to which the effective field theory is
valid.
In other words, we already ``knew" because of eqs. \nine, \ten\  that
the integrals in eqs. \eleven\  could not be taken up to infinity but
up to $\Lambda \leq \Lambda_{triv}$. This introduces a dependence on
$\Lambda$ since the result will depend on the cutoff procedure.
This cutoff dependence on $\Lambda$ may be understood as follows.
When one has accepted the impossibility of removing the cutoff
$\Lambda$, in general there will be higher-dimensional operators
(weighted with the appropriate inverse powers of the cutoff) that are
to be added to the initial Lagrangian. Since the SM is an effective
theory, presumably there will have to be a more fundamental Lagrangian
that sets in at scales $\mu \gsim \Lambda$.
Integrating out these fundamental degrees of freedom above $\Lambda$
is what produces the tower of higher-dimensional operators alluded
to above. In practice, since we do not know what the underlying theory
is for $\mu \gsim \Lambda$, this introduces an uncertainty in the
result in the form of a cutoff dependence.

Now, what is the magnitude of this cutoff dependence? In the limiting
case $g=g'=\lambda=0$ \foot{$g$ and $g'$ are the SM gauge couplings, and
$\lambda$ is the quartic scalar coupling.},
the {\it maximum} cutoff $\Lambda$ is given by $\Lambda_{triv}$ of eq. \ten.
However, not having any information on the
underlying theory for $\mu \gsim \Lambda$, we
cannot know the right way to implement this physical cutoff $\Lambda$. We
estimate this uncertainty by choosing $\Lambda/\Lambda_{triv}=0.6$ and
0.4.
The integrals in eqs. \eleven\  are then
carried out between 0 and $\Lambda$ (instead of $\infty$) and one obtains
for $\tau$ the curves of \fig\figfour{},\
in which we also plot the perturbative result for comparison
(see below). We only plot
in the region $m_t < \Lambda$.
\nl\centerline{\epsfysize=4.5cm\epsfbox{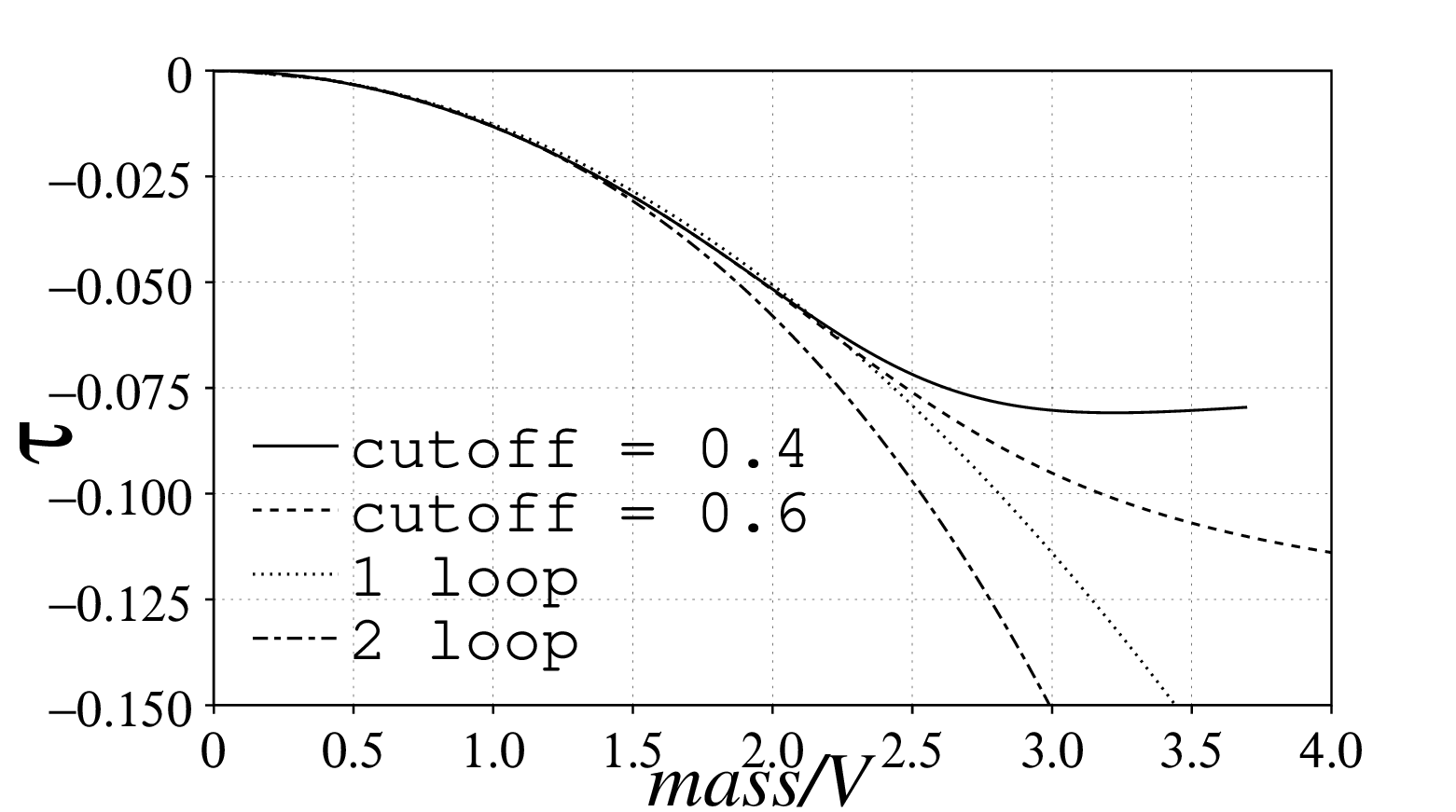}}
\sni\figfour\ {\it $\tau$ versus the top mass to leading order
in the $1/\nc$ expansion for the cutoff values
$\Lambda/\Lambda_{triv}=0.4,0.6$. The two-loop result is extracted from
Barbieri et al. and Fleischer et al. in ref. \twoloop. }

\vskip 0.25in
As one can see, the $1/\nc$ curves tend to saturate, and they deviate from
the perturbative result in the region of large masses, where their behavior
is dominated by cutoff effects.

On the other hand, things go very differently in perturbation theory.
One obtains the perturbative series by expanding eqs. \eleven\ in
$\y^R$. This yields
\eqn\fourteen{\eqalign{\tau(\y^R)=-{\alpha_t \over \nc}-
{1\over \nc}
\sum_{n=0}^{\infty} \alpha_t ^{2+n} 2^n \int_0^{\infty} &{d\xi\over
(1+\xi)^2} \left[
{n-2\over 2} + {n-1\over 2\xi}+ {1-n\over 2} \left(1+{1\over \xi}\right)^2
\log(1+\xi)\right]\cr
&\times \left[\int_0^1 dt\  t \log\left(1+\xi \ (1-t)\right)\right]^n}}
where $\alpha_t\equiv \nc ({\y^R/ 4\pi})^2$.
The first few terms are
\eqn\fifteen{\tau(\y^R)=-{\alpha_t\over \nc} + 0 + {\alpha_t^3\over 4 \nc}
+\ldots}
The $\alpha_t$ term in eq. \fifteen\  of course agrees with the
one-loop result.
The $\alpha_t^2$ term fortuitously vanishes, which means that this
term is of higher order in the $1/\nc$ expansion.
This is in agreement with the calculation of Barbieri et al. in ref.
\zbb\
, who find no $\nc$ dependence in the two-loop result, and explains why the
two-loop curve of fig. 3 tends to deviate from the rest.
The $\alpha_t^3$ term in eq. \fifteen\  can be considered as an
estimate of the three loop contribution.

Higher--order terms can also be computed easily using eq. \fourteen.
For large $n$ one finds the following asymptotic behavior
\eqn\asympt{\tau(\y)\sim {1\over \nc}\ \sum_n \alpha_t^n C_n, \qquad
C_n\sim (n+1)!}
Due to the factorial growth of $C_n$ with $n$, the perturbative
expansion not only has a zero radius of convergence but is also
Borel non-summable. This is the consequence of triviality that may be
gleaned from the large-order behavior of perturbation theory
\ref\RENORMALON{D.J. Gross and A. Neveu, \prd{\bf D10} (1974) 3235;
B. Lautrup, \plb{\bf 69B} (1977) 438;
S. Chadha and P. Olesen, \plb{\bf 72B} (1977) 87;
G. Parisi, \plb{\bf 73B} (1978) 65;
K.~Aoki, \prd{\bf D49} (1994) 1167.}.

We see that no trace of the need for a physical cutoff exists within
perturbation theory, since
by expanding the denominators in eqs. \eleven\
one misses the pole that exists in the loop momentum.
For a small Yukawa coupling, when  the perturbative expansion is
supposed to work, the physical cutoff  given by eq.
\ten\  is fantastically large. Cutoff effects in $\tau$ are
of $\c O(V^2/\Lambda^2_{triv})$, whose precise numerical value
depends on the details of the
cutoff procedure, i.e. on the dynamics of the underlying theory that is
supposed to take over the SM for $\mu>\Lambda_{triv}$.
Notice that the cutoff effects depend on the symmetry breaking scale.
This is why non-decoupling effects are interesting for studying
triviality-related issues. In an ordinary physical process
without non-decoupling effects, it would have
appeared, instead of $V$, the energy scale of the process,
which is typically (much) smaller.

Of course, the $\c O(V^2/\Lambda^2_{triv})$ correction is minute and,
for all practical purposes, completely irrelevant even for a top mass
as large as $250\gev$; this is why one forgets about triviality
altogether when computing radiative
corrections. Moreover, for a top this heavy or lighter, one may
compute higher orders in perturbation theory and add
to the perturbative series in $\y^R$ the radiative corrections coming from
the Higgs boson, for instance. This was first done, at the two-loop level,
by Barbieri et al. in ref.
\twoloop.
The output of this calculation is that the Higgs contributions
eventually grow
with the Higgs mass, clearly manifesting non-decoupling. In perturbation
theory, one is led to believe that as the Higgs becomes heavier, namely in
the range $400 \gev<M_H<1 \tev$, one has to keep adding loop corrections
in order to obtain a more and more precise answer.
However, as the Higgs becomes
heavier, the $\lambda$ coupling becomes larger and its associated
triviality scale goes down exponentially as\foot{This equation is
the result of the combined information extracted from 1/$N$, renormalization
group and lattice techniques
\ref\LKS{L.~Lin, J.~Kuti and Y.~Shen, in  the Proceedings of ``Lattice
Higgs Workshop", Tallahassee (1988) and references therein.}.
It is a trick of Nature
that, for $M_H<1\tev$, it essentially coincides with the result
of naively integrating the one-loop perturbative $\beta$
function for $\lambda$.}
\eqn\aaa{\Lambda^2_{triv}\approx M_H^2 \exp\left({4\pi^2\sqrt{2}\over
3 M_H^2 G_F}\right)\ .}
Clearly, for a sufficiently heavy Higgs (with $M_H < 1\tev$,
nonetheless),  but still $m_t< 200\gev$,
the physical cutoff of the theory will no longer be given in terms of
the  Yukawa, eq. \ten, but in terms of the Higgs, eq. \aaa.
Regretfully our simple $1/\nc$ approach does not allow us to calculate
non-perturbatively
the contribution of the Higgs to the $Z\overline b b$ vertex $\tau$,
because it is a
subleading effect in $1/\nc$.
However our previous calculation with the top
quark shows that we should expect new denominators to appear in eq.
\eleven, of a form similar to that showed by the function $f(\xi)$ of eq.
\thirteen, i.e. producing a pole in eq. \eleven.
We may anticipate that this will be the effect because, as we have
seen in the Yukawa system,  this is the way triviality
modifies perturbation theory, and the Higgs system is also a trivial one.

These new denominators will force us to cutoff the integrals, but now
at the scale
\aaa. We may estimate that
\eqn\aaaa{\tau(m_t,M_H)\sim - {\alpha_t\over \nc}\left(1- \gamma
{m_t^2\over \Lambda_{triv}^2}\right)- {\alpha_t^2\over 2 \nc^2}
\tau^{(2)}(M_H/m_t)  +\ldots}
where the ellipses stand for terms with higher powers of $\alpha_t$
that we can safely neglect since, for $m_t< 200$GeV, $\alpha_t$ is
small.
Here, $\gamma$ is of order 1 and  parametrizes the dependence of
$\tau$ on the precise cutoff procedure, i.e. on the physics beyond the SM.
The $\tau^{(2)}$ term is the two-loop
calculation of Barbieri et al. and of Fleischer et al. in ref. \twoloop .
Rearranging a little, one obtains from eq. \aaaa
\eqn\av{
\tau(m_t,M_H)\sim -{\alpha_t\over \nc} \left\{ 1 + {\alpha_t\over 2 \nc}
\left[\tau^{(2)} - \gamma \left({4\pi V\over M_H}\right)^2
\ \exp\left({-{8\pi^2V^2\over 3M_H^2}}\right)\right]\right\}+\ldots}
The crucial point here is that in a non-perturbative
expression including both the top and the Higgs contributions,
there is no way to separate these contributions, so that
we need to impose an overall cutoff for the whole expression.
In table 1 we present the values  of the cutoff contribution to
$\tau$, which we call $(\Delta \tau)_{triv}$ and which correspond to the
$\gamma$-dependent term in eq. \av. We also show the corresponding
two-loop perturbative contribution, $\tau^{(2)}$, as extracted from
the work of Barbieri et al. and Fleischer et al in ref. \twoloop.
We chose $m_t=200\gev$, or $\alpha_t/\nc=0.0084$.
As
one can see, even for $\gamma\sim1$, $(\Delta \tau)_{triv}$ is about 16\%
of the two-loop perturbative contribution $\tau^{(2)}$  for
$M_H=600\gev$. It can be even higher if $\gamma$ turns out to be two
or three, or if the Higgs is heavier.
\mni\begintable
 $M_H$ | $400\gev$ | $500\gev$ | $600\gev$ | $700\gev$ | $800\gev$ | $900\gev$
\elt
 $\tau^{(2)}$ |1.33| 1.60 | 1.95 | 2.36 | 2.79 | 3.23 \elt
 $\left(\Delta\tau\right)_{triv}$ |0.003| 0.07 | 0.32 | 0.75 | 1.2 | 1.7
\endtable
\catcode`\|=12
\sni Table 1 {\it Comparison of the two-loop contribution to $\tau$
and the ambiguity due to triviality for various Higgs masses.
}%
\subsec{$Zb\overline b$ vertex in the $1/\nf$ expansion}
The large-$\nc$ expansion selects diagrams with closed fermion loops over
other types of topologies because these always have the factor $\nc$
out front.
However it is possible that other types of diagrams turn out to be
numerically more important and that consequently the large-$\nc$
expansion is misleading.
In order to get an idea of how much our conclusions depend on the
precise $1/N$ expansion used, we shall next present the results of a
$1/\nf$ expansion where $\nf$ is now the number of flavors (in the SM,
$\nf=2$).
The large-$\nf$ expansion selects loop diagrams with scalars and
is somewhat complementary to the large-$\nc$ expansion.
It should be straightforward, although much more complicated, to
generalize our result to the
so-called Veneziano limit, where we let both $\nc$ and $\nf$ go to infinity
with the ratio $\nc/\nf$ held fixed \BN.

In order to consider the generalization of the SM to an arbitrary number of
flavors $\nf$, one takes the scalar field to be
$\phi^T=\left({(v+H+i\chi^0)/ \sqrt{2}}, -\chi^1,
-\chi^2,\ldots,-\chi^{\nf-1}\right)$ and the fermion field to be
$\psi_L=(t_L,b^1_L,\ldots,b^{\nf-1}_L)$.
The Lagrangian with only
the scalar fields is the O$(2\nf)$ model
\ref\ON{K.G.~Wilson, \prd{\bf D7} 2911;
L.~Dolan and R.~Jackiw, \prd{\bf D9} (1974) 3320;
H.J.~Schnitzer, \prd{\bf D10} (1974) 1800, 2042;
S.~Coleman, R.~Jackiw and H.D.~Politzer, \prd{\bf D10} (1974) 2491;
  L.F.~Abbott, J.S.~Kang and H.J.~Schnitzer, \prd{\bf D13} (1976) 2212;
  W.A.~Bardeen and M.~Moshe, \prd{\bf D28} (1983) 1372;
M.B.~Einhorn, \npb{\bf B246} (1984) 75; U.M. Heller, H. Neuberger and P.
Vranas, \npb{\bf B399} (1993) 271. }.
The charged Nambu--Goldstone boson may be chosen to be any of the
$\nf-1$ $\chi$'s due to the residual symmetry.
We refer the
reader to \KA\ for details.
We will again compute to lowest order in the weak gauge couplings but
non-perturbatively in the Yukawa coupling.
In this case, unlike the large-$\nc$
case, the
top propagator is corrected at leading order and reads
\eqn\sixteen{\left[i \pslash \left(A_{R,bare}(p^2)\ P_R+P_L\right)+{\y
v\over \sqrt{2}}\right]^{-1},\quad\hbox{where}\quad
A_{R,bare}(p^2)\equiv 1 - {\y^2\nf \over 32\pi^2}\log{p^2\over L^2}\quad ,}
where $L^2$ is a momentum cutoff. Then the Yukawa coupling $\y$ is
renormalized in the following way
\eqn\eighteen{\y^2(\mu)={\y^2(L)\over 1- {\y^2(L) N_F/( 32
\pi^2)} \log{(\mu^2/ L^2)}}\quad ,}
whereas $v$ remains unrenormalized to this order, i.e. $v=V$ in this case.
One can solve eq.
\sixteen\ for the physical mass and width by looking for a pole in the
complex plane  in the form $p^2=(m_t-i \Gamma_t/2)^2$, as a function of
$\y^2(\mu_0)$.
The renormalization scale $\mu_0$ is arbitrary and will not affect any
physical results.

Now we can go about computing $\tau(\y)$. To leading order in $1/\nf$,
the only contribution is the one represented in \fig\figsix{}.
\nl\centerline{\epsfysize=3.0cm\epsfbox{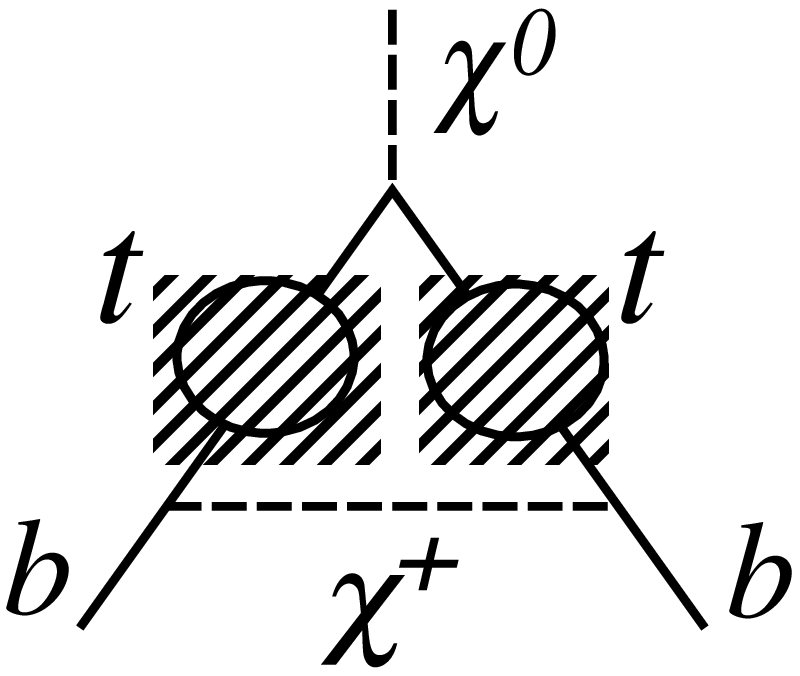}}\smallskip\noindent
\sni\figsix\ {\it Leading-order contributions to  $\tau(\y)$ in the
$1/\nf$ expansion.
Lines with blobs denote the full propagators.}\sni

\vskip 0.25in
A straightforward calculation yields
\eqn\nineteen{\tau(\y)=-{\y^4(\muz) v^2\over 2 (4\pi)^2}\!\!
\int_0^{\Lambda^2} \!\!\!
{dl^2\over \left[l^2 A_R(l^2)+{\y^2(\muz)v^2/
2}\right]^2}\quad\hbox{where}\quad
A_R(l^2)\equiv 1-{\nf\over 2}\left({\y(\muz)\over
4\pi}\right)^2 \log{l^2\over \muz^2}}

Again, this expression for $\tau$ is ill-defined unless we introduce a
cutoff
$\Lambda^2\lesssim\Lambda'^2_{triv}\equiv \muz^2
\exp(32\pi^2/\y(\muz)^2\nf )$, which produces a cutoff dependence of
order $v^2/\Lambda^2$
analogously to the result of the $1/\nc$ expansion.
For a certain arbitrarily chosen $\muz$,
this equation gives $\tau(\y(\muz))$ and since the physical mass is
determined by $\y(\muz)$, one can also obtain $\tau(m_t)$.
Nevertheless,
since the relationship between $\y(\muz)$ and $m_t$ is highly
non-linear it is not
possible to write $\tau(m_t)$ in a simple analytical form and we shall
give the result in the plot of \fig\figseven{}, again for the cutoff
values $\Lambda/\Lambda_{triv}=0.4,0.6$. Clearly, only the region $m_t <
\Lambda$ is of physical significance; however, in this case with a
non-vanishing width, we find it more convenient to use $m_t^2+\Gamma_t^2/4 <
\Lambda^2$\AP , and the $1/\nf$ curves are plotted within this region.
One then sees that, although the details of fig. 3 and fig. 5 are
clearly different, their gross
features are similar. Firstly, one observes clear deviations from the
perturbative contribution only in the region where cutoff effects are also
large. And secondly, although the perturbative curves would grow on forever
with the particle's mass, the $1/N$ curves are bounded in both cases. We
notice that these were also the main characteristics of the non-perturbative
calculation of the $\rho$ parameter done in ref. \AP.

Comparison of $\tau(\y(\muz))$ in eq. \nineteen\  and the result for
the $\rho$ parameter obtained in ref \AP\ leads us to establish the
following amusing non-perturbative relation to leading order in
$1/\nf$:
\eqn\twentyone{\tau(m_t)=-{2\over 3}\left(\rho(m_t)-1\right)\quad .}
\nl\centerline{\epsfysize=4.5cm\epsfbox{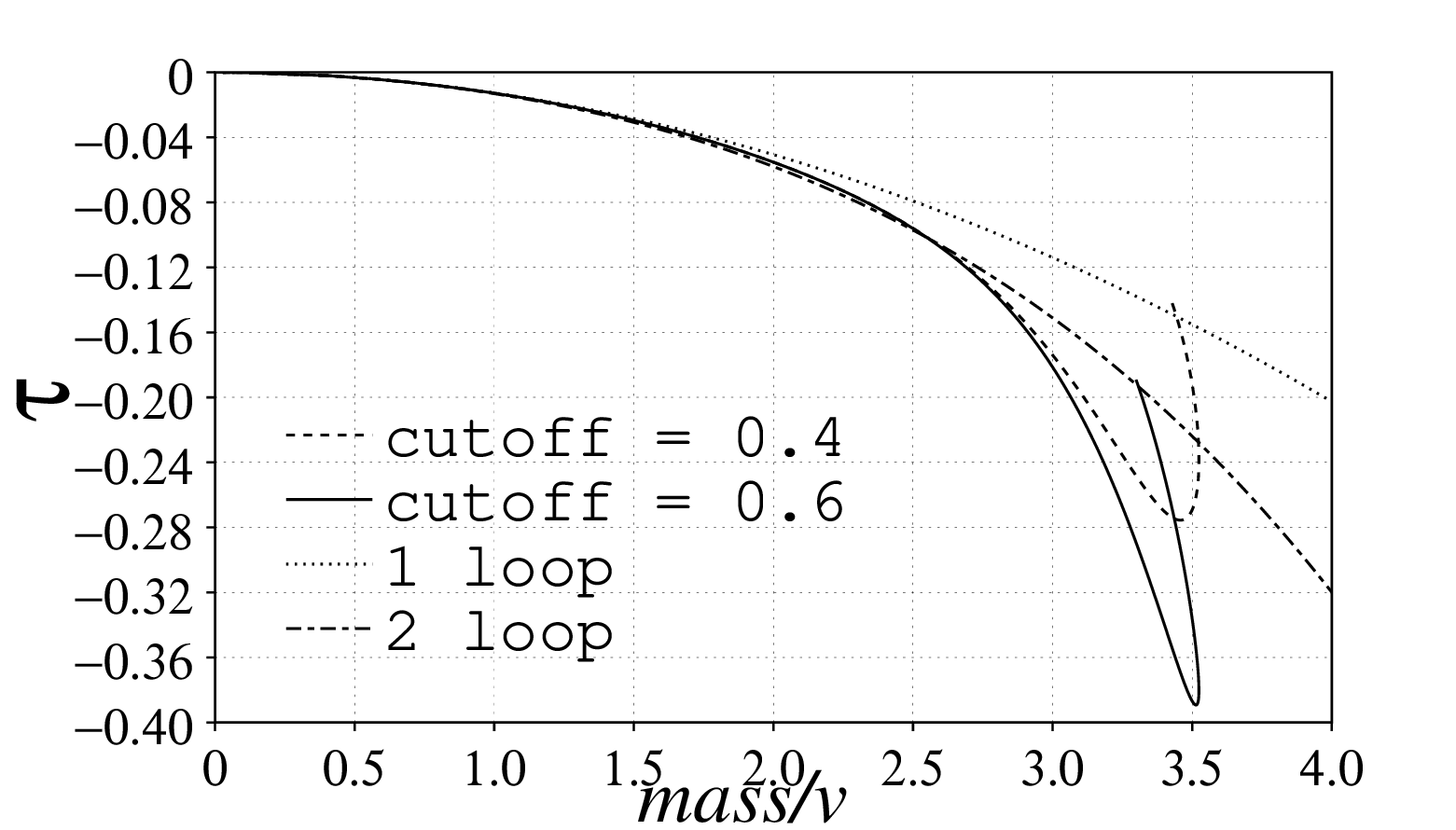}}
\sni\figseven\ {\it $\tau$ plotted against the physical top mass
for the cutoff values $\Lambda/\Lambda_{triv}=0.4,0.6$.}

\vskip 0.25in
In order to compare eq. \nineteen\  to the two-loop perturbative result of ref.
\twoloop , we can also expand eq. \nineteen\  in powers of the top
``mass", defined through the pole of
the real part of the propagator, which is $\y(\mu=m_t)v/\sqrt{2}$, and that
we shall call $\hat m_t$. Clearly this is not the same as the physical mass
(which is defined through the pole of the propagator in the complex plane)
but it suffices for our limited purpose of a two-loop comparison.
The expansion of eq. \nineteen\  in powers of $G_F\hat m_t^2$ yields
\eqn\twentytwo{\tau(\hat m_t)=-{\sqrt{2}G_F\hat m_t^2\over 8 \pi^2}
\left[1+{\nf\sqrt{2}G_F\hat m_t^2\over 16\pi^2}+\ldots\right]_{\nf=2}\quad .}
In this case, the $1/\nf$ result does not have the fortuitous
cancellation of  the $(G_F\hat m^2_t)^2$ term, unlike the
$1/\nc$ expansion.
The perturbative calculations of Barbieri et al. and Fleischer et al.
in ref. \twoloop\ produce the coefficient $(27-\pi^2)/3\simeq5.7$
instead of $\nf$ in eq. \twentytwo,
so that the numerical coefficient turns out to be different, but of
the same order and with the same sign.
The large-order behavior of the perturbative expansion  for
$\tau$ in \nineteen, similar to that of \asympt, is again
consistent with triviality.

Let us consider the case of a heavy Higgs
($400\gev<M_H<1\tev$) when the top is lighter than
$200\gev$.
Again, the role of the Higgs will be to force the cutoff of the
theory to be below the triviality scale due to the top, so that
the effects of new physics may  again be estimated to be of
$\c O(\alpha_t^2v^2/\Lambda^2_{triv})$ as in \aaaa, where
$\Lambda^2_{triv}$ is given by the Higgs coupling \aaa.
The size of this dependence is the same as the one given in table
1.

\subsec{Discussion}

In this work we have used $1/N$ techniques to study how a known
non-perturbative property of the SM, namely triviality, modifies results
obtained in perturbation theory. We took the $Zb\bar b$ vertex as an
example.

{}From figs. 3 and 5 one sees that the most apparent change, with respect to
perturbation theory, is the appearance of cutoff effects. We also see that
the perturbative growth of $\tau$ with the particle's mass, characteristic
of a non-decoupling effect, tends to saturate nonperturbatively and therefore
deviates from the perturbative result, but only in the region where the
cutoff effects are large. These cutoff effects are akin to the scaling
violations found in lattice studies.

There of course remains the open question of how much the SM ($\nf=2,
\nc=3$) resembles the two limits we have taken ($\nf=\infty, \nc=3$ and
$\nf=2, \nc=\infty$). While we cannot claim, and do not claim, that this
resemblance be quantitatively good, we think that it is very reasonable to
expect that the large-$N$ limits we take will not change the gross features
qualitatively. After all, all the large-$N$ limits do is implement triviality
in a consistent way. The cutoff ambiguity and other related effects are just
natural consequences of triviality.

\medskip\noindent{\bf Acknowledgements}

K.A. would like to thank the hospitality of UCLA where a part
of this work was conducted.

S.P. would like to thank Sergio Fanchiotti, Tim Morris and Arcadi Santamaria
for interesting conversations; and Maurizio Consoli, Rolf Tarrach and
Gabriele Veneziano for a critical reading of the manuscript. He would also
like to thank R. Barbieri for a comment explaining the results of ref.
\twoloop. The work of S.P. was partially supported by the research project
CICYT-AEN93-0474.

\listrefs
\end